\documentclass[preprint,amsmath,aps,10pt,superscriptaddress,letterpaper,tightenlines,showpacs]{revtex4}
\usepackage{bm}
\usepackage{stmaryrd}
\usepackage{amsmath}
\usepackage{amssymb}
\usepackage{graphicx}
\usepackage{textcomp}
\usepackage{calrsfs}
\usepackage{yfonts}
\usepackage{color}
\newcommand{\curl}{{\nabla\times}}

\begin{document}
%%%%%%%%%%%%%%%%%%%%%%%%%%%%%%%%%%%%%%%%%%%%%%%%%%%%%%%%%%%%%%%%%%%%%%%%%%%%%%%%%%%%%%%%%%%%%%%%%%%%%%%%%%%%%%%%%%%%%%%%%%%%%%%%%%%%%%%%%%%%%%%%%%%%%%%%%%%%%%%%
\title{Classical and quantum emitters near a metal surface }

\author{Zahra Mohammadi}
\affiliation{Department of Physics,
Faculty of Science, University of Isfahan,
Isfahan, Iran}
\author{Fardin Kheirandish}
\email{fkheirandish@yahoo.com}
\affiliation{Department of Physics,
Faculty of Science, University of Isfahan,
Isfahan, Iran}
%%%%%%%%%%%%%%%%%%%%%%%%%%%%%%%%%%%%%%%%%%%%%%%%%%%%%%%%%%%%%%%%%%%%%%%%%%%%%%%%%%%%%%%%%%%%%%%%%%%%%%%%%%%%%%%%%%%%%%%%%%%%%%%%%%%%%%%%%%%%%%%%%%%%%%%%%%%%%%%%
\begin{abstract}
In the present article the SPP propagation in an attenuating medium is investigated. The analytical calculations of
the total electric-field Green's tensor of a metal-dielectric interface structure are provided and novel explicit expressions for Green's tensor
of a metal-dielectric interface are presented. The contribution of plasmons are obtained by evaluating the poles of the reflection coefficient for p-polarized waves incident on the metal interface. The emission pattern of a classical dipole located above air/silver interface is studied.
The relative intensity of the field to the field intensity in free space is studied for both normal and parallel orientations of the dipole. The quantum optical properties of a quantum emitter coupled to a metal surface are studied. For a quantum dot near a metal surface single photon emission is demonstrated using second-order correlation functions.
\end{abstract}
\maketitle
%%%%%%%%%%%%%%%%%%%%%%%%%%%%%%%%%%%%%%%%%%%%%%%%%%%%%%%%%%%%%%%%%%%%%%%%%%%%%%%%%%%%%%%%%%%%%%%%%%%%%%%%%%%%%%%%%%%%%%%%%%%%%%%%%%%%%%%%%%%%%%%%%%%%%%%%%%%%%%%%
\section{Introduction}
%%%%%%%%%%%%%%%%%%%%%%%%%%%%%%%%%%%%%%%%%%%%%%%%%%%%%%%%%%%%%%%%%%%%%%%%%%%%%%%%%%%%%%%%%%%%%%%%%%%%%%%%%%%%%%%%%%%%%%%%%%%%%%%%%%%%%%%%%%%%%%%%%%%%%%%%%%%%%%%%
The surface plasmon polaritons (SPP) are EM waves that are resulted due to the interaction of light and a metallic surface \cite{Rivera}.
They are of interest to a wide spectrum of scientists, ranging from physicists, chemists and materials scientists. In special, the desire
to control and manipulate light at nanoscale have renewed the interest in surface plasmons \cite{Archambault1, Barnes}.
Two level systems, quantum dots (QD) and electric dipole antennas located close to a metal nanostructure are basic components for quantum plasmonics \cite{Novotny}.
Moreover, nanoplasmonics is a very active field of study that have attracted a great deal of attention in recent years \cite{Novotny, Nanotechno}.

In a previous paper, we analysed a planar perfect surface as an important limiting case of a perfect conductor wedge \cite{Mohammadi1}.
We know that plasmonic excitation is not possible for the perfect conductors. The results for a real metal differ significantly from
that of a perfect electric conductor since for an ideal metal surface the only available enhancement mechanism is scattering. Having this motivation, we have considered a real metal surface to demonstrate strong enhancement of light near a plasmonic surface.

In the present work we discuss the emission of radiation by localized systems of oscillating charge and current densities. We have considered the radiation from a dipole antenna or a sinusoidally oscillating charge positioned sufficiently near a real metal surface. The plasmonic nanofocusing structures produce a strong enhancement and confinement of a local field. We have also obtained and plotted the plasmonic electric field enhancements of the dipole antenna system as a function of its distance to a metal surface \cite{Mirzaee}.
Our theoretical results may lead to new optical manipulation methods for nanoscale optical communication, biophotonics, nanoscale lithography and
medical testing \cite{Rivera, antenna1, antenna2, Jia}. Also it is expected that many new applications will be developed for such physical systems in the years to come.

The response of a dipole placed in front of a complex structure can be characterized using the Green's function approach \cite{Kheirandish1, Kheirandish2, Kheirandish3}.
So, we will present a useful and applicable expression for Green's tensor of a metal-dielectric interface structure, where the plasmonic contribution $D_{spp}$ will be obtained by extracting the contribution of the pole of the reflection coefficient for p-polarized waves
incident on the metal interface  \cite{Archambault2}. Since the normal orientation is the optimal direction to couple with the
metal surface, a majority of works have considered only perpendicular polarization, here we have considered both orientations.
We shown that for the case of a dipole pointing perpendicular to the metal surface, the strong electric field enhancement
can be resulted in contrast to the case of parallel orientation of the dipole that is in agreement with the results reported in \cite{Mirzaee, Novotny, Gonzalez2014}.

Also, for developing nanophotonic single photon devices, a study on the second-order correlation property of the
fluorescence is necessary and important \cite{G2.2016}. Second-order correlation function is a typical nonclassical property of
light, which demonstrates the single photon emission \cite{Harouni, Gonzalez2010, G2.2016, G2.2010}. Single-photon sources play a central role in
light-based quantum-information systems and in modern quantum optical applications. The demonstration of efficient, scalable, on-chip single photon sources is
one of the most important challenges \cite{single photon}. In the final section, we have investigated the second-order
correlation property of the the light emission from a QD placed near a planar metal.

The paper is arranged as follows: In Sec. II, the basic theory is reviewed. In sec. III, Green's tensor for a metal-dielectric
interface structure and the method to construct the corresponding Green's tensor $D_{spp}$ are presented. In sec. IV, we apply the results of the
preceding sections and study enhancement of SPP field around a dipole antenna near a metal surface. In Sec. V, we have considered a QD near a metal surface, and the single photon emission is demonstrated using second-order correlation
measurement. Finally, we conclude in sec VI.
%%%%%%%%%%%%%%%%%%%%%%%%%%%%%%%%%%%%%%%%%%%%%%%%%%%%%%%%%%%%%%%%%%%%%%%%%%%%%%%%%%%%%%%%%%%%%%%%%%%%%%%%%%%%%%%%%%%%%%%%%%%%%%%%%%%%%%%%%%%%%%%%%%%%%%%%%%%%%%%%
\begin{figure}
    \includegraphics[scale=0.35]{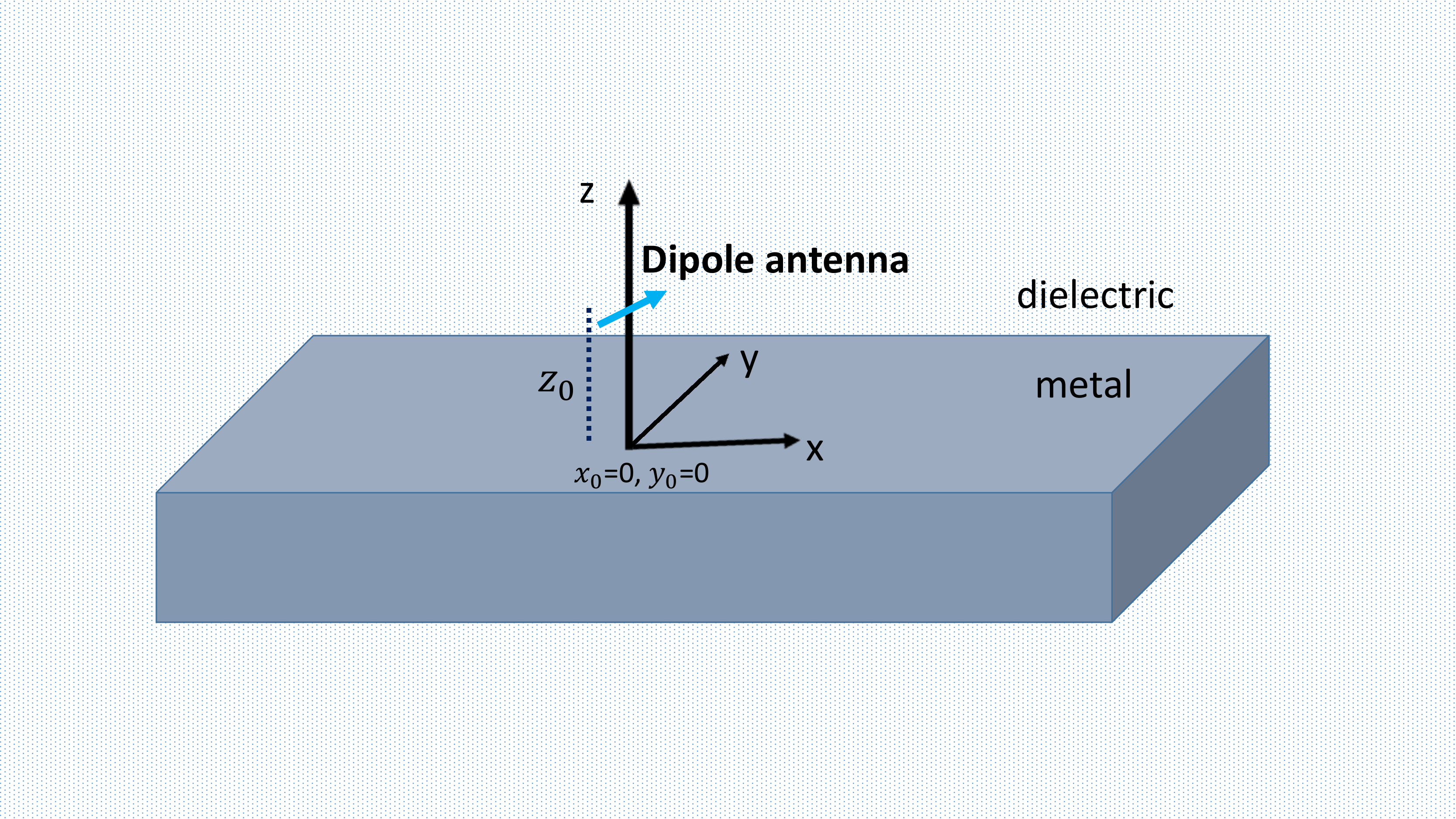}\\
  \caption{(Color online) A dipole antenna located above a dielectric-metal interface.
}\label{Fig1}
\end{figure}
%%%%%%%%%%%%%%%%%%%%%%%%%%%%%%%%%%%%%%%%%%%%%%%%%%%%%%%%%%%%%%%%%%%%%%%%%%%%%%%%%%%%%%%%%%%%%%%%%%%%%%%%%%%%%%%%%%%%%%%%%%%%%%%%%%%%%%%%%%%%%%%%%%%%%%%%%%%%%%%%
\section{Basic formulae}
%%%%%%%%%%%%%%%%%%%%%%%%%%%%%%%%%%%%%%%%%%%%%%%%%%%%%%%%%%%%%%%%%%%%%%%%%%%%%%%%%%%%%%%%%%%%%%%%%%%%%%%%%%%%%%%%%%%%%%%%%%%%%%%%%%%%%%%%%%%%%%%%%%%%%%%%%%%%%%%%
\subsection{second order correlation function}
%%%%%%%%%%%%%%%%%%%%%%%%%%%%%%%%%%%%%%%%%%%%%%%%%%%%%%%%%%%%%%%%%%%%%%%%%%%%%%%%%%%%%%%%%%%%%%%%%%%%%%%%%%%%%%%%%%%%%%%%%%%%%%%%%%%%%%%%%%%%%%%%%%%%%%%%%%%%%%%%
To investigate the changes in the quantum statistics of a QD in the presence of an interface, we consider a QD
placed above a planar metal surface. The second order coherence function in the resonant case can be analytically obtained as \cite{Gonzalez2010, G2.2016}
\begin{equation}\label{g2}
g^{(2)}(\tau)=1-e^{-\frac{3\Gamma\tau}{4}}\,\big[\cos(R\tau)+\frac{3\Gamma}{4R}\sin(R\tau)\big],
\end{equation}
where
\begin{equation}
R=\sqrt{\Omega^2-\frac{\Gamma^2}{16}},
\end{equation}
is Rabi splitting at resonance and the decay rate is given by
\begin{equation}\label{decay}
  \Gamma=\frac{2}{\hbar}\omega^{2}\mbox{Im}[\mathbf{d}_0\cdot\mathbf{D}(\mathbf{r}_{0},
  \mathbf{r}_{0},\omega)\cdot\mathbf{d}_0],
\end{equation}
where $\mathbf{d}_0$ is the dipole moment of QD. One can find that $ g^{(2)}(\tau) > g^{(2)}(0)$, which is a typical nonclassical
property of light, i.e. the anti-bunching character \cite{G2.2016}. To calculate the second-order correlation from Eq.(\ref{g2}), we need to find the corresponding dyadic Green function $\mathbf{D}$. In the following, we will study the Green's tensor of a planar interface.
%%%%%%%%%%%%%%%%%%%%%%%%%%%%%%%%%%%%%%%%%%%%%%%%%%%%%%%%%%%%%%%%%%%%%%%%%%%%%%%%%%%%%%%%%%%%%%%%%%%%%%%%%%%%%%%%%%%%%%%%%%%%%%%%%%%%%%%%%%%%%%%%%%%%%%%%%%%%%%%%
\begin{figure}
    \includegraphics[scale=0.35]{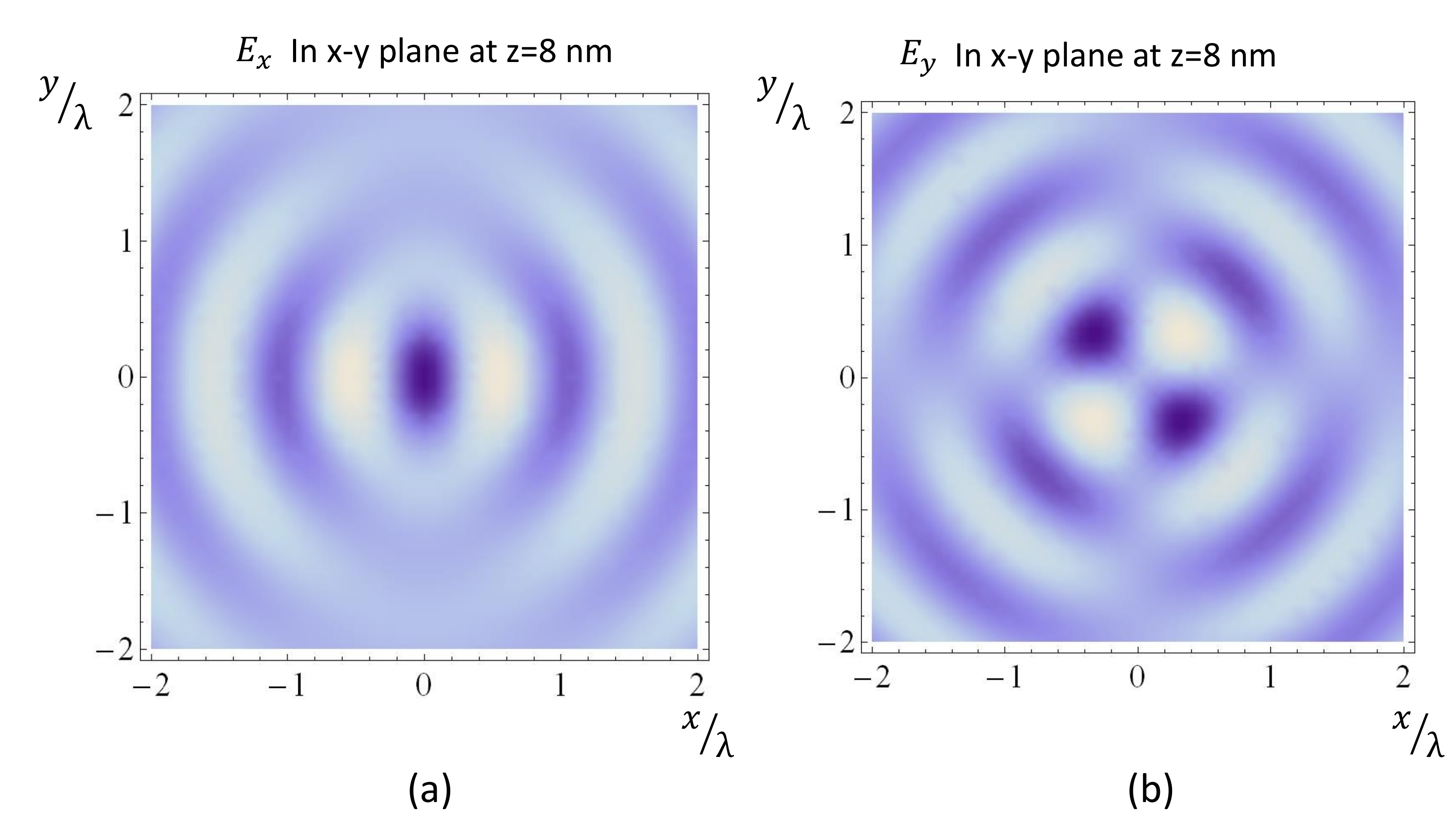}\\
  \caption{(Color online) Components of electric field in x-y plane for a dipole in x direction. $(\lambda= 500nm, l=20 nm, z_{0}=10nm) $
}\label{Fig2}
\end{figure}
%%%%%%%%%%%%%%%%%%%%%%%%%%%%%%%%%%%%%%%%%%%%%%%%%%%%%%%%%%%%%%%%%%%%%%%%%%%%%%%%%%%%%%%%%%%%%%%%%%%%%%%%%%%%%%%%%%%%%%%%%%%%%%%%%%%%%%%%%%%%%%%%%%%%%%%%%%%%%%%%
\begin{figure}
    \includegraphics[scale=0.4]{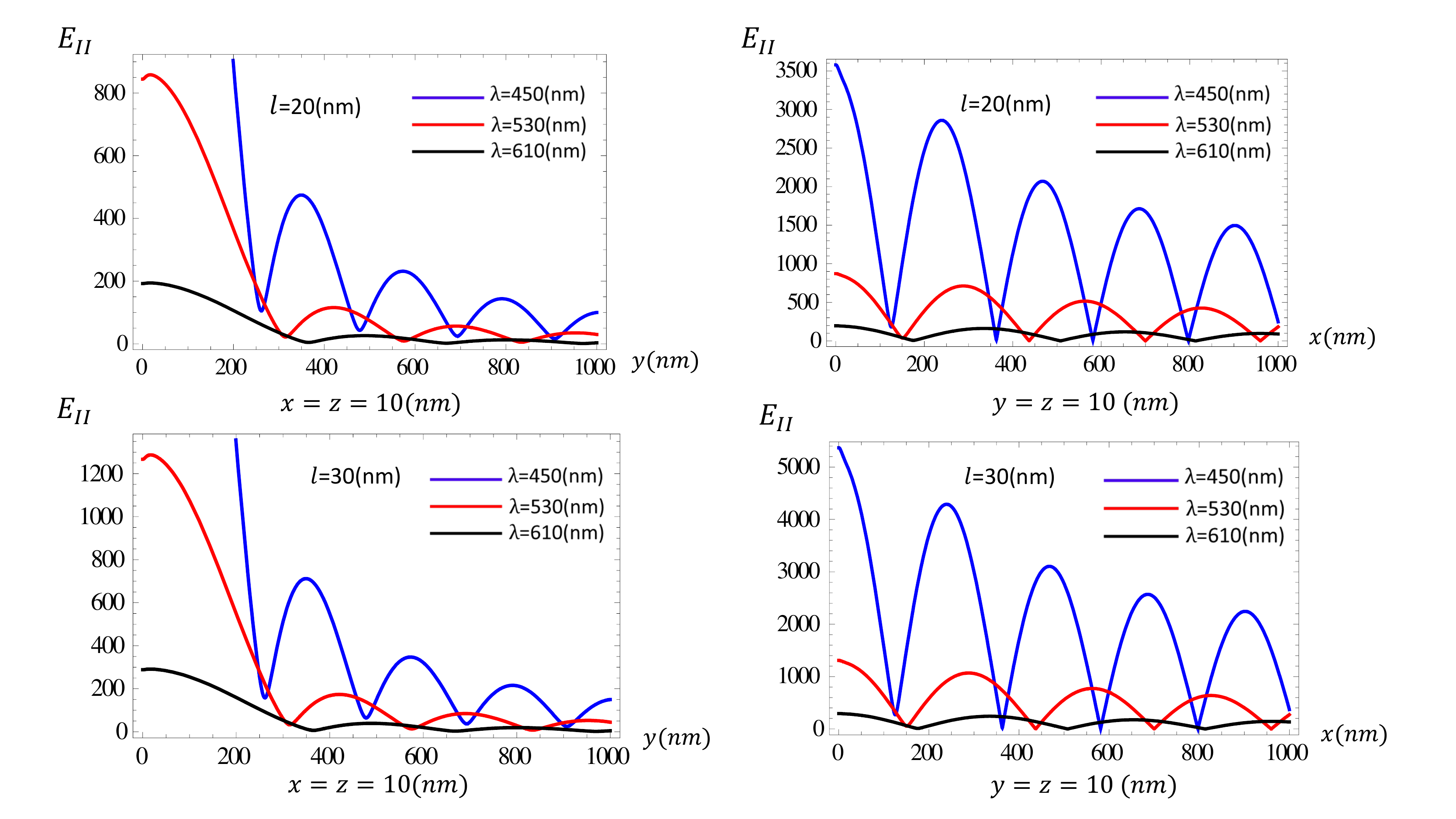}\\
  \caption{(Color online) $E_{\|}$ as a function of x (Left Panels) and y (Right Panels).
  The dipole is oriented along x axes with different lengths.}\label{Fig3}
\end{figure}
%%%%%%%%%%%%%%%%%%%%%%%%%%%%%%%%%%%%%%%%%%%%%%%%%%%%%%%%%%%%%%%%%%%%%%%%%%%%%%%%%%%%%%%%%%%%%%%%%%%%%%%%%%%%%%%%%%%%%%%%%%%%%%%%%%%%%%%%%%%%%%%%%%%%%%%%%%%%%%%%
\subsection{Effective products for the electric field}
From Maxwell's equations, the electric field $\mathbf{E}(\mathbf{r}; \omega)$ in a general, linear, isotropic magnetodielectric medium,
satisfies the wave equation \cite{Kheirandish3}
\begin{equation}\label{general}
 \curl(\frac{\mathbf{1}}{\mu}\curl\mathbf{E})-\frac{\omega^2}{c^2}\,\epsilon\mathbf{E} =
\mu_0\omega^2\mathbf{P}^{N}+i\mu_0\omega\curl\mathbf{M}^{N},
\end{equation}
where $\mathbf{P}^{N}(\mathbf{r},\omega)$ and $\mathbf{M}^{N} (\mathbf{r},\omega)$ are polarization and magnetization noise fields and
$\epsilon(r,\omega)$ and $\mu(r,\omega)$ are dimensionless permittivity and permeability of the medium, respectively. The constant $\mu_0$
is the permeability of the vacuum. Here we assume for simplicity a nonmagnetic medium. In the presence of an external current density $\mathbf{J}(\mathbf{r},\omega)$ and outside the material medium we have
\begin{equation}\label{general}
 \big[\curl\curl-\frac{\omega^2}{c^2}\,\epsilon(\mathbf{r},\omega)\big]\,\mathbf{E}(\mathbf{r},\omega) =i\omega\mu_{0}\,\mathbf{J}(\mathbf{r},\omega).
\end{equation}
The geometry considered here is depicted in Fig.1, where the dipole is embedded in the half space $z > 0$ having real permittivity $\epsilon_{1}(\omega)=1$.
The medium in the half space $z < 0 $ is characterized by a complex permittivity $\epsilon_{2}(\omega)$. In Drude model, the permittivity $\epsilon_{2}(\omega)$ is given by
\begin{equation}\label{permittivity}
\epsilon_{2}(\omega)=\epsilon_{\infty}\,(1-\frac{\omega_{p}^2}{\omega(\omega+i\gamma_{p})}).
\end{equation}
For silver, in the range of frequencies of interest, the related parameters are: plasma frequency ($\omega_{p}=3.76\,ev$)
, high-frequency limit ($\varepsilon_{\infty}=9.6$), and damping constant ($\gamma_{p}=0.03\,\omega_{p}$) \cite{Gonzalez2010}.
Eq. (\ref{general}) is an inhomogeneous wave equation and the solution can be expressed in terms of the retarded dyadic Green's function $G(\mathbf{r}; \mathbf{r'}; \omega)$, a rank two tensor, which is the solution to the Helmholtz equation
\begin{equation}\label{green}
 [\curl\curl-\frac{\omega^2}{c^2}\,\epsilon(\mathbf{r},\omega)]\mathbf{G}(\mathbf{r}; \mathbf{r'}; \omega) =\mathbf{1}\delta(\mathbf{r}-\mathbf{r}').
\end{equation}
The electric field in real space generated by a source current $\mathbf{j}(\mathbf{r}, t)$ can be determined using Fourier inverse transform and Green's function as \cite{Matloob}
\begin{equation}\label{E}
\mathbf{E}(\mathbf{r},t)=\frac{\mu_{0}i}{\sqrt{2\pi}}\,\int\omega\, d\omega \,e^{-i\omega t}\,\int d\mathbf{r} \, \mathbf{G}(\mathbf{r},\mathbf{r}',\omega)\,\cdot\,\mathbf{j}(\mathbf{r}',\omega).
\end{equation}
%%%%%%%%%%%%%%%%%%%%%%%%%%%%%%%%%%%%%%%%%%%%%%%%%%%%%%%%%%%%%%%%%%%%%%%%%%%%%%%%%%%%%%%%%%%%%%%%%%%%%%%%%%%%%%%%%%%%%%%%%%%%%%%%%%%%%%%%%%%%%%%%%%%%%%%%%%%%%%%%
\begin{figure}
    \includegraphics[scale=0.4]{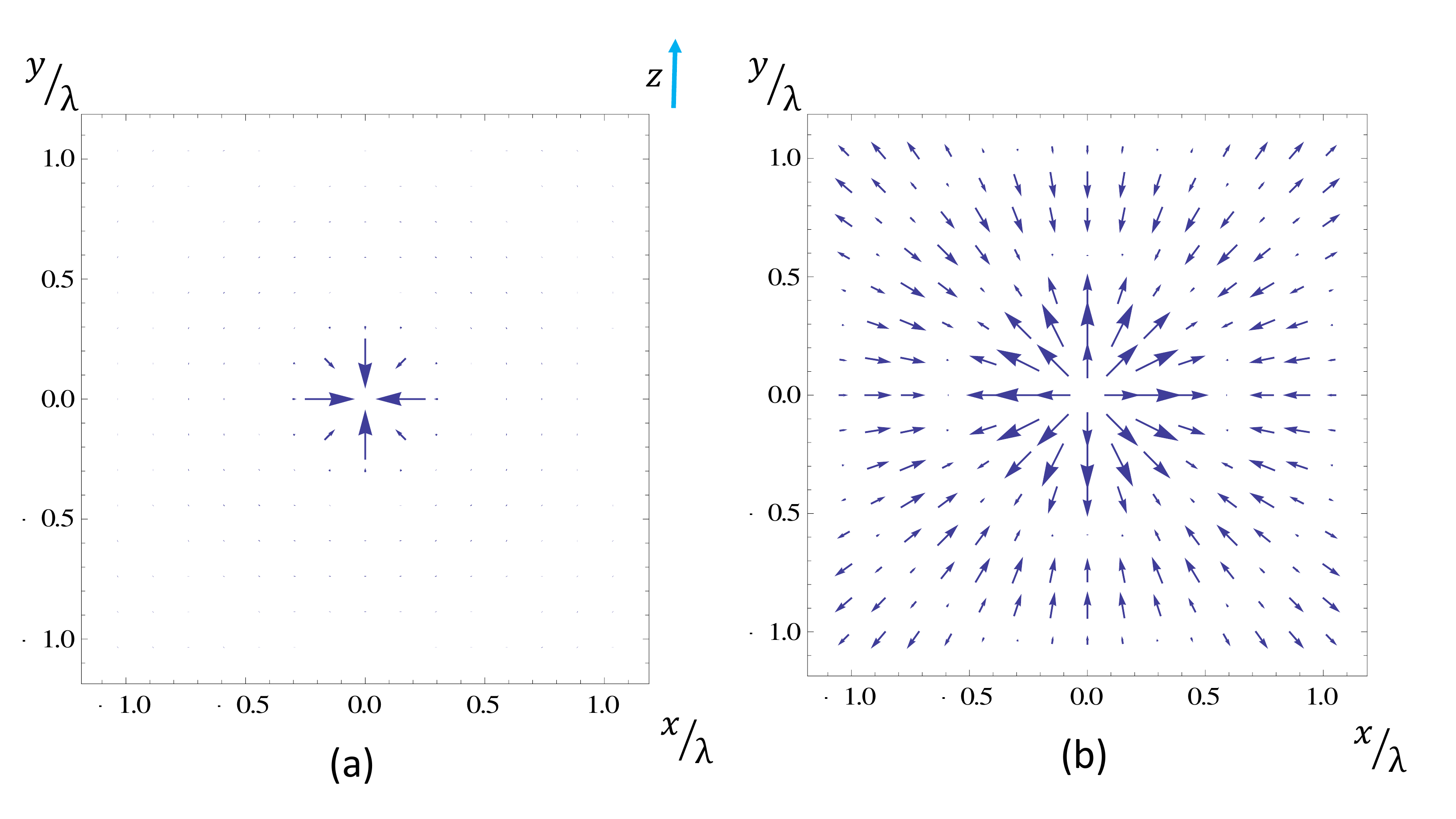}\\
  \caption{(Color online) Field distribution around a perpendicular dipole at distance $z_0=30$ nm  above the air/silver interface. The figure shows
  the surface plasmon propagation along the top surface of silver.
 Arrows: direction and magnitude of the electric field. ($\lambda= 580\,$ nm, $l=20$ nm, $z=100\,$ nm) }\label{Fig4}

\end{figure}
%%%%%%%%%%%%%%%%%%%%%%%%%%%%%%%%%%%%%%%%%%%%%%%%%%%%%%%%%%%%%%%%%%%%%%%%%%%%%%%%%%%%%%%%%%%%%%%%%%%%%%%%%%%%%%%%%%%%%%%%%%%%%%%%%%%%%%%%%%%%%%%%%%%%%%%%%%%%%%%%
\begin{figure}
    \includegraphics[scale=0.4]{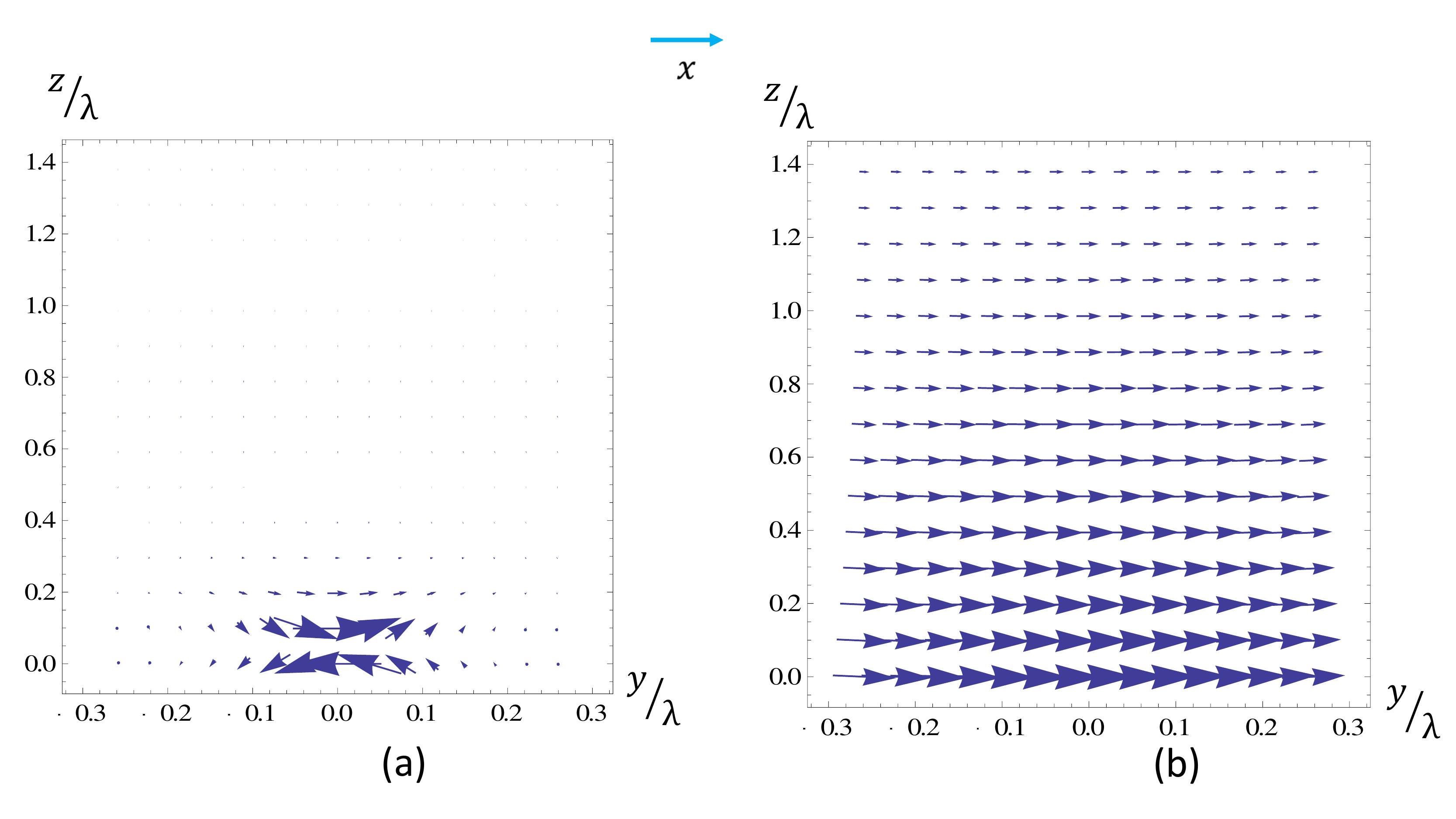}\\
  \caption{(Color online) Field distribution around a parallel dipole at a distance of $z_0=30$ nm  above the air/silver interface. The figure shows
  the surface plasmon propagation along the top surface of silver.
 Arrows: direction and magnitude of the electric field. ($\lambda= 580\,$nm, $l=20\,$ nm, $x=50\,$nm) }\label{Fig5}
\end{figure}
%%%%%%%%%%%%%%%%%%%%%%%%%%%%%%%%%%%%%%%%%%%%%%%%%%%%%%%%%%%%%%%%%%%%%%%%%%%%%%%%%%%%%%%%%%%%%%%%%%%%%%%%%%%%%%%%%%%%%%%%%%%%%%%%%%%%%%%%%%%%%%%%%%%%%%%%%%%%%%%%
\section{Green's function of dielectric-metal interface}
%%%%%%%%%%%%%%%%%%%%%%%%%%%%%%%%%%%%%%%%%%%%%%%%%%%%%%%%%%%%%%%%%%%%%%%%%%%%%%%%%%%%%%%%%%%%%%%%%%%%%%%%%%%%%%%%%%%%%%%%%%%%%%%%%%%%%%%%%%%%%%%%%%%%%%%%%%%%%%%%
In this section we will construct Green's tensor of dielectric-metal interface. This dyadic Green's function can be calculated following approaches presented in reference \cite{Tai1, Tai2, Milton, Prachi, Maradudin}. Here we apply the method introduced in \cite{Milton, Prachi, Maradudin} where the whole problem is reduced to finding scalar Green's functions subject to boundary conditions imposed by boundary conditions (see Appendix A). In this section, we use the results obtained in Appendix A and focus on the surface wave contribution defined as the pole contribution. Also we will obtain the term $D_{spp}$ from the total Green's tensor by extracting the contribution to the pole of the reflection coefficient for p-polarized waves incident on the metal interface.

From Appendix A, it is seen that the components of matrix $\mathbf{g}$ has poles given by the denominator of the Fresnel
factors for p-polarized field. For a dielectric/metal interface, they correspond to the surface plasmon as discussed
previously. We evaluate the pole contribution to the Green's tensor by integrating in the complex $k$-plane and denote $k_{spp}$ as a root of the denominator \cite{Sondergaard}.
In some works emphasis is placed on the fields in the metal or dielectric \cite{Berman}. Here, the dipole as a source, is placed at a distance $z_{0}$ above a metal surface and we will show the enhancement of the field in the vicinity (above) of metallic surface. Thus, we just need the Green's function in the upper half-space ($z >0 and z'>0 $). We consider both orientations of the dipole, normal and parallel, so we need the whole dyadic Green tensor components. For example, let us find $D_{zz,spp}(K, z, z′, ω)$.

From Eqs. (\ref{Fourier}, \ref{dbold} , \ref{gzz}), we find
\begin{eqnarray}\label{D_{zz}}
  \mathbf{D^{z,z'>0}_{zz,spp}}(\mathbf{X},\mathbf{X}';\omega)&=&\frac{ i\, c^2}{2\pi\,\omega^2  \epsilon^d}
  \int \frac{\mathbf{k}^{2}_{\shortparallel}\,d^2 \mathbf{k}_{\shortparallel}}{k_d}\,e^{i \mathbf{k}_{\shortparallel}
  \cdot(\mathbf{x}_{\shortparallel}-\mathbf{x}'_{\shortparallel})}\,\frac{k_m \epsilon^d +k_d \epsilon^m}{k_m \epsilon^d -k_d \epsilon^m}
e^{i k_d (z+z')},\nonumber\\
&=&\frac{ i\, c^2}{2\pi\,\omega^2  \epsilon^d}\int\limits_{0}^{+\infty} \frac{k^{3}_{\shortparallel}\,d k_{\shortparallel}}
{\,k_d}\,\int\limits_{0}^{2\pi} d\theta\,e^{i (k_{\shortparallel}\cos\theta\cdot(x-x')+k_{\shortparallel}\sin\theta\cdot(y-y'))}
\,\frac{k_m \epsilon^d +k_d \epsilon^m}{k_m \epsilon^d -k_d \epsilon^m}
e^{i k_d (z+z')},
\end{eqnarray}
where we made use of the identity
\begin{equation}
\int\limits_{0}^{2\pi} d\theta\,e^{i (k_{\shortparallel}\cos\theta\cdot(x-x')+k_{\shortparallel}\sin\theta\cdot(y-y'))}=2\pi\,J_{0}(k_{\shortparallel}\,\sqrt{(x-x')^2+(y-y')^2}).
\end{equation}
Using Taylor expansion of the denominator around the roots up to the first order of approximation and doing some algebra, the Green's tensor can be obtained as
\begin{equation}\label{Dtensor}
  \mathbf{D^{z,z'>0}_{zz,spp}}(\mathbf{X},\mathbf{X}';\omega)=\frac{ i\, c^2}{\omega^2  \epsilon^d}\int\limits_{0}^{+\infty} \frac{k^{3}_{\shortparallel}\,d k_{\shortparallel}}{\,k_{\shortparallel}}\,\frac{k_m \epsilon^d +k_d \epsilon^m}{(k_{\shortparallel}-k_{spp})}\,(\frac{-\epsilon_{m}}{\epsilon^{2}_{d}-\epsilon^{2}_{m}})\,J_{0}(k_{\shortparallel}\,\sqrt{(x-x')^2+(y-y')^2})
e^{i k_d (z+z')}.
\end{equation}
From calculus of residues, one can evaluate the integral in Eq. \ref{Dtensor}
\begin{equation}
  \mathbf{D^{z,z'>0}_{zz,spp}}(\mathbf{X},\mathbf{X}';\omega)=-2\pi i\,|k_{spp}|\,\frac{\epsilon^{3}_{m}\,\epsilon_{d}}{\sqrt{\epsilon_{d}(-\epsilon_{m})}\,(\epsilon_{d}+\epsilon_{m})\,(\epsilon^{2}_{d}-\epsilon^{2}_{m})}
  \,J_{0}(k_{spp}\,\sqrt{(x-x')^2+(y-y')^2})\,e^{-\sqrt{\frac{\epsilon_{d}}{-\epsilon_{m}}}\,|k_{spp}|\,(z+z')}.
\end{equation}
Now using Eqs. (\ref{green},\ref{greenD}) and $D_{spp}(\mathbf{X},\mathbf{X}';\omega)= 4\pi\,G_{spp}(\mathbf{X},\mathbf{X}';\omega)$, we will find
\begin{equation}\label{GZZ}
\mathbf{G^{z,z'>0}_{zz,spp}}(\mathbf{X},\mathbf{X}';\omega)=\frac{-i|k_{spp}|}{2}
\frac{\varepsilon_{d}\varepsilon_{m}^3}{\sqrt{\varepsilon_{d}(-\varepsilon_{m})}(\varepsilon_{d}+\varepsilon_{m})(\varepsilon_{d}^2-\varepsilon_{m}^2)}
J_{0}(|k_{spp}|\mid \vec{r}_{\|}-\vec{r}'_{\|}\mid)\,e^{-\sqrt{\frac{\varepsilon_{_{d}}}{-\varepsilon_{m}}}|k_{spp}|(z+z')}.
\end{equation}
In a similar way the calculation of all of the remaining components of Green's function can be done. The results of these calculations
are summarized and the explicit form of components of Green's tensor $D_{spp}(\mathbf{X},\mathbf{X}';\omega)$ are given in Appendix A.
%%%%%%%%%%%%%%%%%%%%%%%%%%%%%%%%%%%%%%%%%%%%%%%%%%%%%%%%%%%%%%%%%%%%%%%%%%%%%%%%%%%%%%%%%%%%%%%%%%%%%%%%%%%%%%%%%%%%%%%%%%%%%%%%%%%%%%%%%%%%%%%%%%%%%%%%%%%%%%%%
\begin{figure}
    \includegraphics[scale=0.4]{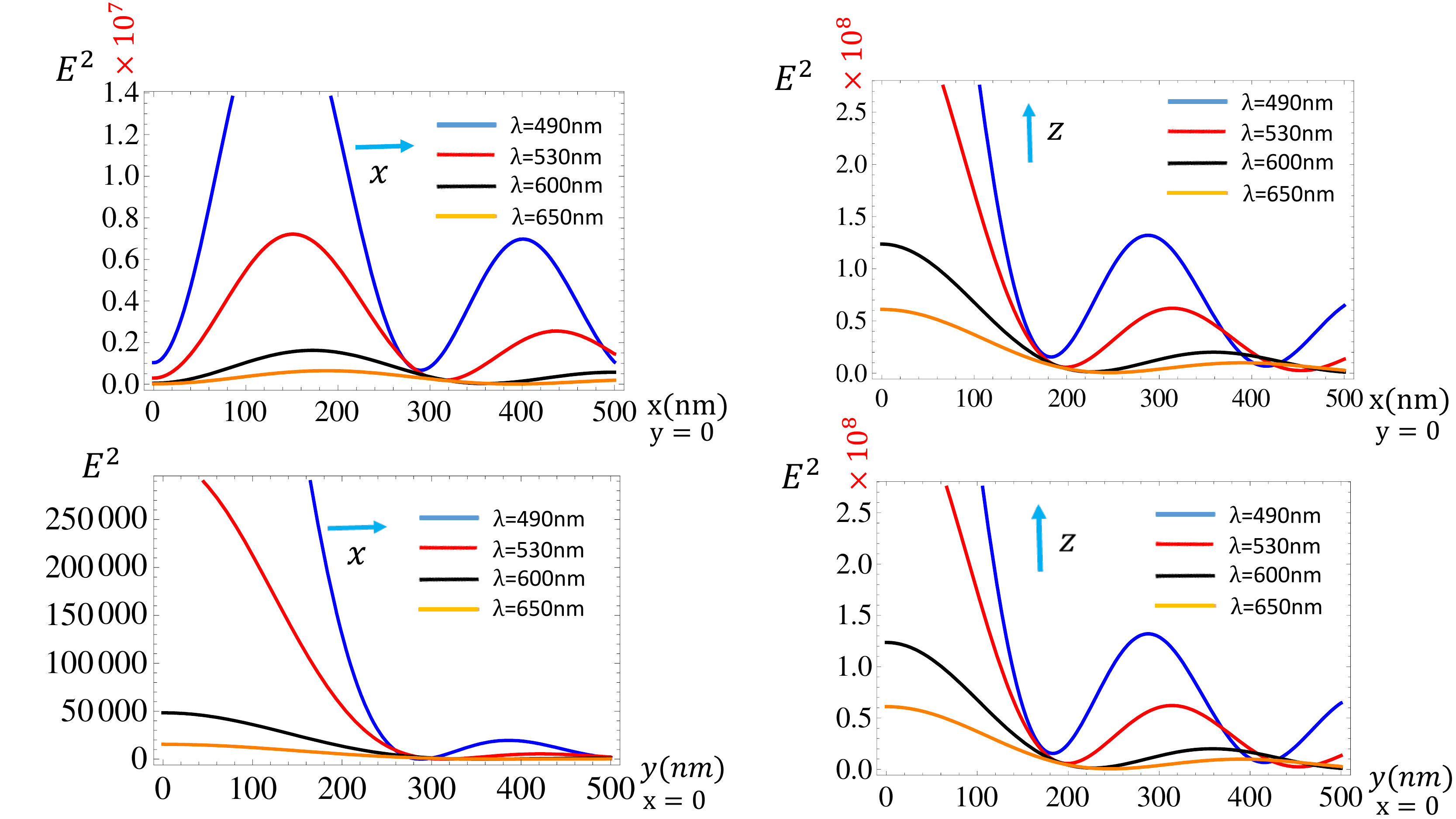}\\
  \caption{(Color online) Field enhancement near a metal surface at $z_{0}=10 nm$ as a function of radial distance on x-y plane.
   Different plots correspond to the different
values of the wavelengths. (l=15nm, z=8nm)}\label{Fig6}
\end{figure}
%%%%%%%%%%%%%%%%%%%%%%%%%%%%%%%%%%%%%%%%%%%%%%%%%%%%%%%%%%%%%%%%%%%%%%%%%%%%%%%%%%%%%%%%%%%%%%%%%%%%%%%%%%%%%%%%%%%%%%%%%%%%%%%%%%%%%%%%%%%%%%%%%%%%%%%%%%%%%%%%
\begin{figure}
    \includegraphics[scale=0.4]{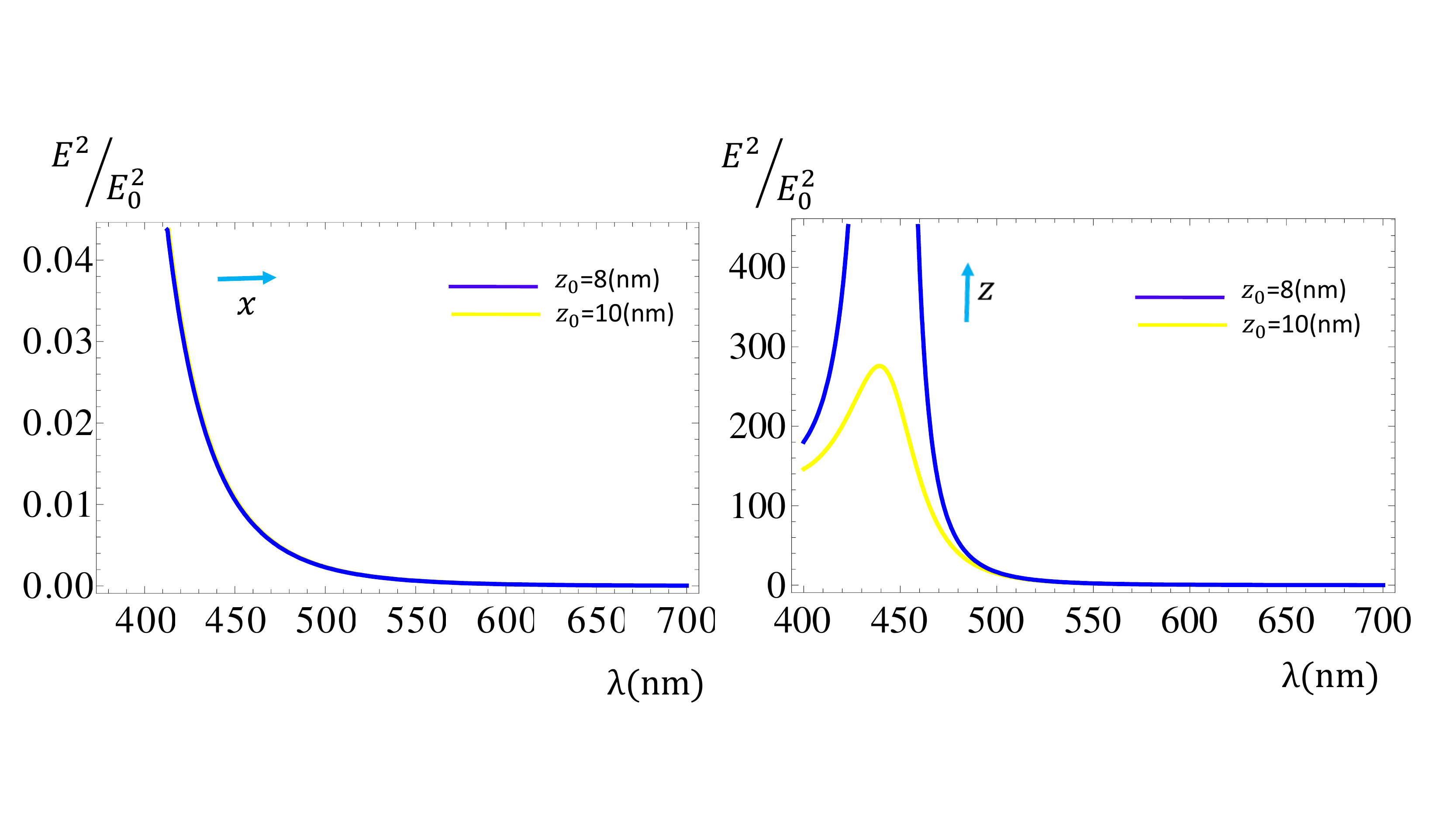}\\
  \caption{(Color online) Modification of electric field for different heights of the dipole $z_{0}$ above
  the dielectric-metal interface as
a function of wavelength. (l=15nm, x=50nm, y=50nm, z=10nm).}\label{Fig7}
\end{figure}
%%%%%%%%%%%%%%%%%%%%%%%%%%%%%%%%%%%%%%%%%%%%%%%%%%%%%%%%%%%%%%%%%%%%%%%%%%%%%%%%%%%%%%%%%%%%%%%%%%%%%%%%%%%%%%%%%%%%%%%%%%%%%%%%%%%%%%%%%%%%%%%%%%%%%%%%%%%%%%%%
\section{Surface plasmon field generated by a classical dipole}
%%%%%%%%%%%%%%%%%%%%%%%%%%%%%%%%%%%%%%%%%%%%%%%%%%%%%%%%%%%%%%%%%%%%%%%%%%%%%%%%%%%%%%%%%%%%%%%%%%%%%%%%%%%%%%%%%%%%%%%%%%%%%%%%%%%%%%%%%%%%%%%%%%%%%%%%%%%%%%%%
In this section, we apply the results of the preceding sections and study modification of field components for a point-dipole at different distances to the metal surface. We consider a dipole antenna positioned sufficiently near the metal surface. As an example of such
a system we consider a localized oscillating charge or a dipole antenna of length $l$ and frequency $\Omega$.
The antenna is assumed to be at distance $z_{0}$ above the metal surface oriented along the z and x axes and
we choose a coordinate system with origin on the interface  as indicated in Fig.1. The $z$-coordinate of the dipole ($zo$) denotes the height
of the dipole above the metal surface. The current density can be written as
\begin{eqnarray}
  \vec{j}(\vec{r},t)=\frac{ql\Omega}{2}\,\sin(\Omega t)\delta(y)\delta(x)\delta(z-z_{0})\hat{u},
\end{eqnarray}
where $u$ denotes $x$ or $z$ directions. Equivalently, the Fourier component of the current density is defined by
\begin{eqnarray}\label{jw}
  \vec{j}(\vec{r},\omega)=i\sqrt{\frac{\pi}{2}}\,\frac{ql\Omega}{2}\,\delta(y)\delta(x)\delta(z-z_{0})\,(\delta(\omega-\Omega)-\delta(\omega+\Omega))\hat{u}.
\end{eqnarray}
For calculating the different components of the electric field, we use the Green's tensor $D_{spp}(K, z, z′, ω)$ (see Appendix A). Then
insertion of Eq. (\ref{jw}) into Eq. (\ref{E}) provides the electric field $ E_{spp}$ which is given in Appendix D.

In Fig.2, it is shown that for a dipole in $x$-direction, the symmetry of plasmonic field in $xy$-plane is broken so that the component $ E_{x}$ is stronger at $x$-direction while the component $ E_{y}$ is symmetric in $xy$-plane. This is also illustrated in Fig. 3. As we see, the parallel component to the surface $E_{\|}$ is stronger in the direction that the dipole points.

Fig.4 shows the components of the electric field ($E_{x}$ and $E_{y}$) for a normal dipole emitting in free space (Fig.4(a)) and near a silver surface (Fig.4(b)). Comparison of the two graphs illustrates that for a dipole in free space, the electric field exists only in a very near region of the dipole while when the dipole is placed near the metallic surface the plasmonic field is considerably enhanced. Fig. 5 shows analogous results for a dipole oriented parallel to the surface. Again we see that for a dipole in free space the electric field is significant only in the region around the dipole (Fig.5(a)),
while for a dipole near the surface, the plasmonic field is enhanced near the surface and this field becomes weaker as the distance from the surface
is increased as expected (Fig.5(b)) .

In Fig. 6, the intensity $E^2$ as a function of the distance from the dipole to the $xy$-plane is depicted for different wavelengths.
A clear decay in intensity is observed as the distance increases, and this damping of intensity is lower for larger wavelengths \cite{Van}.
As we know, metal's dielectric constant depends on external electromagnetic field. For silver in the range of frequencies of interest, the imaginary
part of the permittivity is positive, then the SPP modes will be attenuated \cite{Gonzalez2010, Allameh}. This damping is caused by ohmic loss
property of the metal participating in the SPP field \cite{Novotny, Allameh}.

In Figs. (7,8) the relative intensity defined as the field intensity normalized to the field intensity in free space ($\frac{E^{2}}{E_{0}^{2}}$) is depicted in  wavelength for different distances to the surface.

As is illustrated in Fig.7, at short distances away from the dipole on the $xy$-plane ($x=50$nm, $y=50$nm),
the enhancement of electric field does not occur for parallel orientation (Fig. 7(a)) while for the normal orientation,
(Fig. 7(b)), the observed enhancement can be considerable. In Fig. 8, we see that for farther regions in the $xy$-plane around (x=300nm, y=300nm) for both orientations of the dipole, the relative intensity $\frac{E^{2}}{E_{0}^{2}}$ is increasing. This is because for a dipole in free space, as we saw in
Figs. (4, 5), the electric field is considerable only at short distances around the dipole. As is shown in Fig. 8(a), for parallel orientation of the dipole, the enhancement is significant only for farther distances on the $xy$-surface parallel with the plasmonic surface, however Fig. 8(b) shows a strong enhancement for the perpendicular dipole near the interface. Note that for a parallel dipole, the field enhancement is less than the perpendicular case \cite{Novotny, Mirzaee, Gonzalez2014}. This result is in agreement with the idea that in general the perpendicular dipoles exhibit larger enhancement around the metal surfaces \cite{Mirzaee, Le Ru}. It should be noted that for both orientations of the dipole, the normalized intensity is enhanced when reducing the distance between the dipole and the surface, as expected.
%%%%%%%%%%%%%%%%%%%%%%%%%%%%%%%%%%%%%%%%%%%%%%%%%%%%%%%%%%%%%%%%%%%%%%%%%%%%%%%%%%%%%%%%%%%%%%%%%%%%%%%%%%%%%%%%%%%%%%%%%%%%%%%%%%%%%%%%%%%%%%%%%%%%%%%%%%%%%%%%
\begin{figure}
    \includegraphics[scale=0.4]{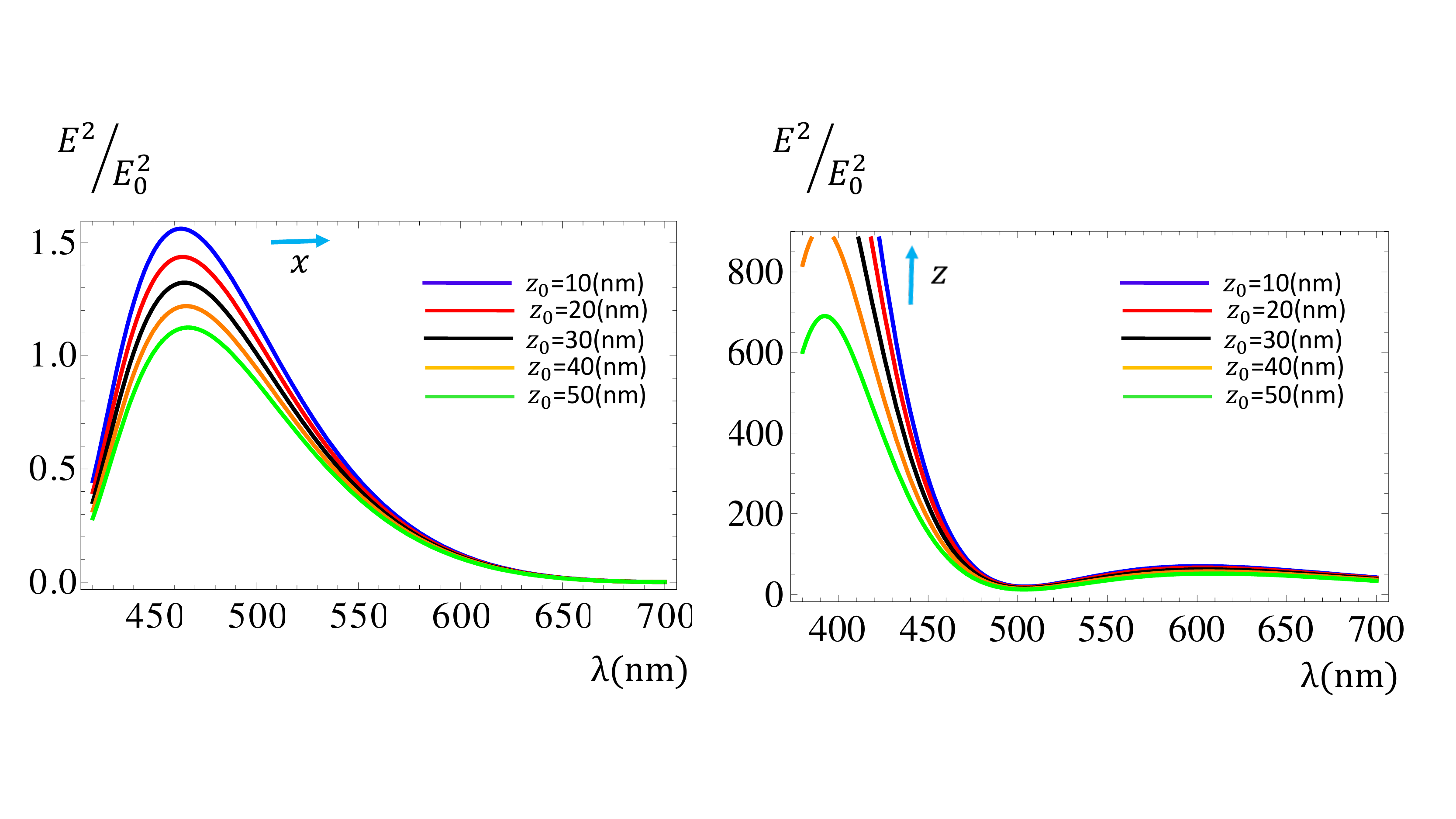}\\
  \caption{(Color online) Modification of the electric field resulted from different heights of the dipole $z_{0}$ above
  the dielectric-metal interface as
a function of wavelength. (l=15nm, x=300nm, y=300nm, z=8nm)}\label{Fig8}
\end{figure}
%%%%%%%%%%%%%%%%%%%%%%%%%%%%%%%%%%%%%%%%%%%%%%%%%%%%%%%%%%%%%%%%%%%%%%%%%%%%%%%%%%%%%%%%%%%%%%%%%%%%%%%%%%%%%%%%%%%%%%%%%%%%%%%%%%%%%%%%%%%%%%%%%%%%%%%%%%%%%%%%
\section{second order correlation function}
%%%%%%%%%%%%%%%%%%%%%%%%%%%%%%%%%%%%%%%%%%%%%%%%%%%%%%%%%%%%%%%%%%%%%%%%%%%%%%%%%%%%%%%%%%%%%%%%%%%%%%%%%%%%%%%%%%%%%%%%%%%%%%%%%%%%%%%%%%%%%%%%%%%%%%%%%%%%%%%%
The correlation properties and possible realizations of single-photon sources are the topic of this section. Using Eq.(\ref{g2}), second-order correlation function $g^{2}(\tau)$, for a QD at distance $z_{0}$ above the silver surface is shown in Fig. 9. This figure exhibits the anti-bunched nature of the emitted photons for both polarizations (normal and parallel) of QD with different values of the wavelength of incident light. According to this figure, the incident field with a larger wavelength leads to a better SP (single photon) emission. Also, we will illustrate the difference between the behavior of two polarizations (normal and perpendicular) of the QD. Then, we can conclude that when the orientation of QD dipole is along the z-direction there is a better SP emission. In addition, in Fig. 10, we observe $g^{2}(\tau)$ for two QD dipole orientations at different distances above the silver surface. As we will see, by decreasing the distance of QD to the surface, the quality of SP emission is increased. The figures indicate that one can control the single-photon emission of the QD by changing spatial orientation, wavelength of incident light on QD, and the distance to the surface \cite{Harouni, G2.2016}.
%%%%%%%%%%%%%%%%%%%%%%%%%%%%%%%%%%%%%%%%%%%%%%%%%%%%%%%%%%%%%%%%%%%%%%%%%%%%%%%%%%%%%%%%%%%%%%%%%%%%%%%%%%%%%%%%%%%%%%%%%%%%%%%%%%%%%%%%%%%%%%%%%%%%%%%%%%%%%%%%
\begin{figure}
    \includegraphics[scale=0.35]{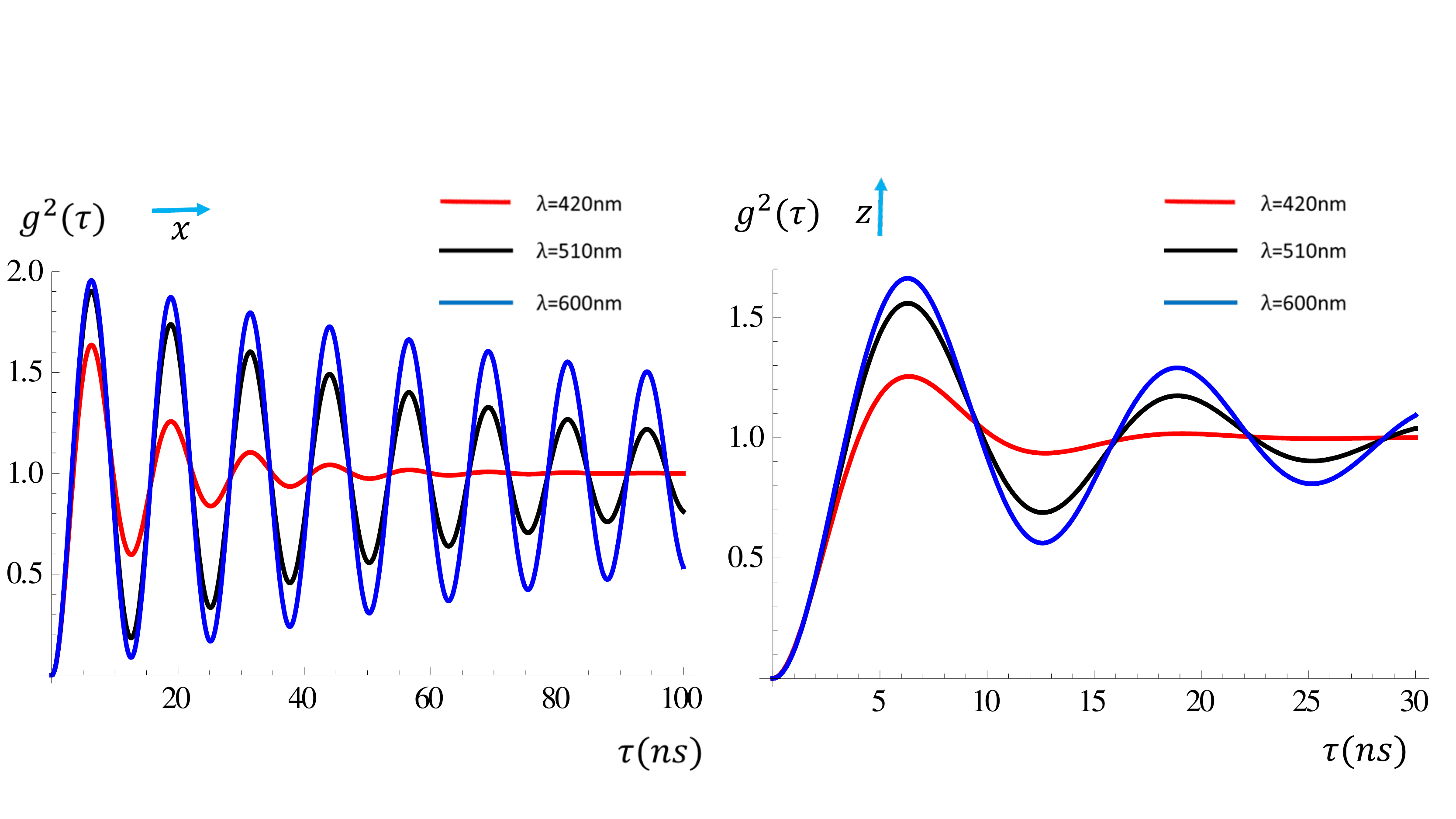}\\
  \caption{(Color online) Second-order correlation function of the emitted photons from a QD with
   dipole direction in the x-axis and z-axis, for different values of the wavelength.}\label{Fig9}
\end{figure}
%%%%%%%%%%%%%%%%%%%%%%%%%%%%%%%%%%%%%%%%%%%%%%%%%%%%%%%%%%%%%%%%%%%%%%%%%%%%%%%%%%%%%%%%%%%%%%%%%%%%%%%%%%%%%%%%%%%%%%%%%%%%%%%%%%%%%%%%%%%%%%%%%%%%%%%%%%%%%%%%
\begin{figure}
    \includegraphics[scale=0.35]{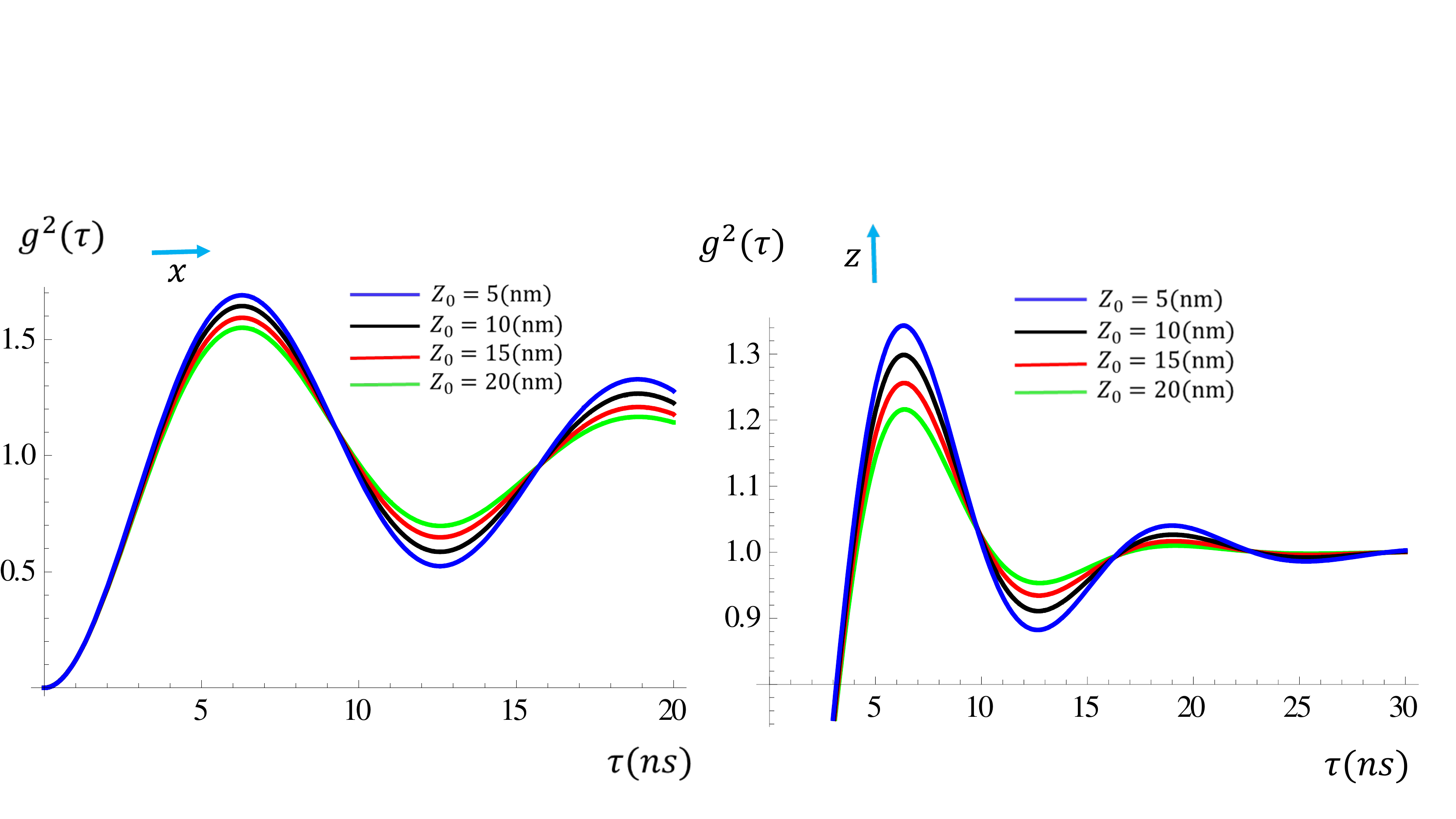}\\
  \caption{(Color online) Second-order correlation function of the emitted photons from a QD with
   dipole direction along the x and z axis for different values of the distance to the air/silver interface.}\label{Fig10}
\end{figure}
%%%%%%%%%%%%%%%%%%%%%%%%%%%%%%%%%%%%%%%%%%%%%%%%%%%%%%%%%%%%%%%%%%%%%%%%%%%%%%%%%%%%%%%%%%%%%%%%%%%%%%%%%%%%%%%%%%%%%%%%%%%%%%%%%%%%%%%%%%%%%%%%%%%%%%%%%%%%%%%%
\section{Conclusions}
%%%%%%%%%%%%%%%%%%%%%%%%%%%%%%%%%%%%%%%%%%%%%%%%%%%%%%%%%%%%%%%%%%%%%%%%%%%%%%%%%%%%%%%%%%%%%%%%%%%%%%%%%%%%%%%%%%%%%%%%%%%%%%%%%%%%%%%%%%%%%%%%%%%%%%%%%%%%%%%%
In the present work, several issues regarding surface plasmons on flat surfaces are discussed using dyadic Green's function approach.
First the surface plasmon field generated by a classical dipole is studied. The results show that the field modification pattern near an air/silver interface differ at different orientations and distances from the surface. Then, the surface plasmonic effect modification to effectively modulate single quantum dots is studied and the single photon emission is demonstrated using second-order correlation function. For both polarizations (normal and parallel), the results reveal the anti-bunching nature of emitted photons from a QD. Our findings also indicate that one can control the single-photon emission of a QD by changing spatial orientation, wavelength of incident light on QD and the distance to the surface.
%%%%%%%%%%%%%%%%%%%%%%%%%%%%%%%%%%%%%%%%%%%%%%%%%%%%%%%%%%%%%%%%%%%%%%%%%%%%%%%%%%%%%%%%%%%%%%%%%%%%%%%%%%%%%%%%%%%%%%%%%%%%%%%%%%%%%%%%%%%%%%%%%%%%%%%%%%%%%%%%
\appendix
\section{}
%%%%%%%%%%%%%%%%%%%%%%%%%%%%%%%%%%%%%%%%%%%%%%%%%%%%%%%%%%%%%%%%%%%%%%%%%%%%%%%%%%%%%%%%%%%%%%%%%%%%%%%%%%%%%%%%%%%%%%%%%%%%%%%%%%%%%%%%%%%%%%%%%%%%%%%%%%%%%%%%
Using a Fourier transform in $x$-$y$ directions, we can define the dimensionally reduced dyadic \cite{Maradudin, Milton, Prachi}
\begin{equation}\label{Fourier}
  \mathbf{D}(\mathbf{X},\mathbf{X}';\omega)=\int \frac{d^2 \mathbf{k}_{\shortparallel}}{(2\pi)^2}\,e^{i \mathbf{k}_{\shortparallel}\cdot
  (\mathbf{X}_{\shortparallel}-\mathbf{X}'_{\shortparallel})}\,\mathbf{d}(\mathbf{k}_{\shortparallel},\omega|z,z')
\end{equation}
where $\mathbf{k}_{\shortparallel}$ and $\mathbf{X}_{\shortparallel}$ are two-dimensional vectors given
by $(k_{x}, k_{y}, 0)$ and $(x, y, 0)$, respectively. The function $ \mathbf{D}(\mathbf{X},\mathbf{X}';\omega)$ is the Green's function satisfying the equation
\begin{equation}\label{greenD}
 [\curl\curl-\frac{\omega^2}{c^2}\,\epsilon(\mathbf{X},\omega)]\mathbf{D}(\mathbf{X},\mathbf{X}';\omega) =4\pi\,\mathbf{1}\delta(\mathbf{X}-\mathbf{X}').
\end{equation}
Due to the rotational symmetry in the $x$-$y$ directions, we can use the matrix S($\mathbf{k}_{\shortparallel}$) as the matrix which
rotates the vector $(k_{x}, k_{y}, 0)$  into the vector ($\mathbf{k}_{\shortparallel}$, 0, 0)
\begin{equation}\label{s}
  \mathbf{s} (k_{\shortparallel})=\frac{1}{k_{\shortparallel}}\left(
                                             \begin{array}{ccc}
                                               k_x & k_y & 0 \\
                                               -k_y & k_x & 0 \\
                                               0 & 0 & k_{\shortparallel} \\
                                             \end{array}
                                           \right),\,\,\,\,\mathbf{s}^{-1} (k_{\shortparallel})=\frac{1}{k_{\shortparallel}}\left(
                                             \begin{array}{ccc}
                                               k_x & -k_y & 0 \\
                                               k_y & k_x & 0 \\
                                               0 & 0 & k_{\shortparallel} \\
                                             \end{array}
                                           \right).
\end{equation}
Now we introduce the matrix $\mathbf{g}$ which is related to the matrix $\mathbf{d}$ by
\begin{equation}\label{dg}
  \mathbf{d}=\mathbf{s}^{-1}(k_{\shortparallel})\,\mathbf{g}\,\mathbf{s} (k_{\shortparallel}),
\end{equation}
therefore
\begin{equation}\label{gbold}
\mathbf{g}=\left(
             \begin{array}{ccc}
               g_{xx} & 0 & g_{xz} \\
               0 & g_{yy} & 0 \\
               g_{zx} & 0 & g_{zz} \\
             \end{array}
           \right),
\end{equation}
and
\begin{equation}\label{dbold}
  \mathbf{d}=\frac{1}{k^2_{\shortparallel}}\left(
               \begin{array}{ccc}
                 k^2_x g_{xx}+k^2_y g_{yy} & k_x k_y (g_{xx}-g_{yy}) & k_x k_{\shortparallel} \\
                 k_x k_y (g_{xx}-g_{yy}) & k^2_x g_{yy}+k^2_y g_{xx} & g_{xz} k_y k_{\shortparallel} \\
                 k_x k_{\shortparallel} g_{zx} & k_y k_{\shortparallel} g_{zx} & k^2_{\shortparallel} g_{zz} \\
               \end{array}
             \right).
\end{equation}
Substitution of Eq.(\ref{Fourier}) into Eq. (\ref{green}) and using Eqs. (\ref{dbold}, \ref{gbold}, \ref{dg}) yields a
set of differential equations for the components of $g$ matrix that can be solved using boundary conditions.
The details of our calculations are not mentioned here and the final results are summarized as
\begin{equation}\label{gxx}
g_{xx}=\left\{
       \begin{array}{ll}
         -\frac{2\pi i k_m c^2}{\omega^2 \epsilon^m}\big[\frac{k_m \epsilon^d +k_d \epsilon^m}{k_m \epsilon^d -k_d \epsilon^m}
\,e^{i k_m (z+z')}-e^{-i k_m |z-z'|}\big], & z<0,\,\,z'<0 \\
        -\frac{4\pi i c^2}{\omega^2}\frac{k_d k_m}{k_m \epsilon^d -k_d \epsilon^m}\,e^{i k_d z+i k_m z'}, & z>0,\,\,z'<0 \\
-\frac{4\pi i c^2 k_d}{\omega^2 \epsilon^d}\frac{k_m}{k_m \epsilon^d -k_d \epsilon^m}\,e^{i k_m z+i k_d z'}, & z<0,\,\,z'>0 \\
    -\frac{2\pi i k_d c^2}{\omega^2 \epsilon^d}\big[\frac{k_m \epsilon^d +k_d \epsilon^m}{k_m \epsilon^d -k_d \epsilon^m}
\,e^{i k_d (z+z')}+e^{i k_d |z-z'|}\big], & z>0,\,\,z'>0
       \end{array}
     \right.
\end{equation}
\begin{equation}\label{gyy}
 g_{yy}=\left\{
    \begin{array}{ll}
      \frac{2\pi i}{k_m}\big[\frac{k_m+k_d}{k_m-k_d}\,e^{i k_m (z+z')}+e^{-i k_m |z-z'|}\big], & z<0,\,\, z'<0 \\
      \frac{4\pi i}{k_m-k_d}\,e^{i k_d z+i k_m z'}, & z>0,\,\, z'<0 \\
      \frac{4\pi i}{k_m-k_d}\,e^{i k_m z+i k_d z'}, & z<0,\,\,z'>0 \\
      \frac{2\pi i}{k_d}\big[\frac{k_m+k_d}{k_m-k_d}\,e^{i k_d (z+z')}-e^{i k_d |z-z'|}\big], & z>0,\,\, z'>0
    \end{array}
  \right.
\end{equation}
\begin{equation}\label{gzz}
g_{zz}=\left\{
         \begin{array}{ll}
           \frac{2\pi i k^2_{\shortparallel} c^2}{\omega^2 k_m \epsilon^m}\big[\frac{k_m \epsilon^d +k_d \epsilon^m}{k_m \epsilon^d -k_d \epsilon^m}
e^{i k_m (z+z')}-e^{i k_m |z-z'|}\big]+\frac{4\pi c^2}{\omega^2 \epsilon^m}\,\delta(z-z'), & z<0,\,\, z'<0 \\
           \frac{4\pi i k^2_{\shortparallel} c^2}{\omega^2}\frac{1}{k_m \epsilon^d -k_d \epsilon^m}\,e^{i k_d z+i k_m z'}, & z>0,\,\, z'<0 \\
           \frac{4\pi i k^2_{\shortparallel} c^2}{\omega^2}\frac{1}{k_m \epsilon^d -k_d \epsilon^m}\,e^{i k_m z+i k_d z'}, & z<0,\,\,z'>0 \\
           \frac{2\pi i k^2_{\shortparallel} c^2}{\omega^2 k_d \epsilon^d}\big[\frac{k_m \epsilon^d +k_d \epsilon^m}{k_m \epsilon^d -k_d \epsilon^m}
e^{i k_d (z+z')}-e^{i k_d |z-z'|}\big]+\frac{4\pi c^2}{\omega^2 \epsilon^d}\,\delta(z-z'), & z>0,\,\, z'>0
         \end{array}
       \right.
\end{equation}
\begin{equation}\label{gxz}
  g_{xz}=\left\{
           \begin{array}{ll}
             -\frac{2\pi i k_{\shortparallel} c^2}{\omega^2 \epsilon^m}\big[\frac{k_m \epsilon^d +k_d \epsilon^m}{k_m \epsilon^d -k_d \epsilon^m}
\,e^{i k_m (z+z')}-e^{-i k_m |z-z'|}\,sgn(z-z'), & z<0,\,\, z'<0 \\
             -\frac{4\pi i k_{\shortparallel} c^2}{\omega^2}\frac{k_d}{k_m \epsilon^d -k_d \epsilon^m}\,e^{i k_d z+i k_m z'}, & z>0,\,\, z'<0 \\
             -\frac{4\pi i k_{\shortparallel} c^2}{\omega^2}\frac{k_m}{k_m \epsilon^d -k_d \epsilon^m}\,e^{i k_m z+i k_d z'}, & z<0,\,\,z'>0 \\
             -\frac{2\pi i k_{\shortparallel} c^2}{\omega^2 \epsilon^d}\big[\frac{k_m \epsilon^d +k_d \epsilon^m}{k_m \epsilon^d -k_d \epsilon^m}
\,e^{i k_d (z+z')}-e^{i k_d |z-z'|}\,sgn(z-z'), & z>0,\,\, z'>0
           \end{array}
         \right.
\end{equation}
\begin{equation}\label{gzx}
  g_{zx}=\left\{
           \begin{array}{ll}
             \frac{2\pi i k_{\shortparallel} c^2}{\omega^2 \epsilon^m}\big[\frac{k_m \epsilon^d +k_d \epsilon^m}{k_m \epsilon^d -k_d \epsilon^m}\,
e^{i k_m (z+z')}+e^{-i k_m |z-z'|}\,sgn(z-z')\big], & z<0,\,\, z'<0 \\
             \frac{4\pi i k_{\shortparallel} c^2}{\omega^2}\frac{k_m}{k_m \epsilon^d -k_d \epsilon^m}\,e^{i k_d z+i k_m z'}, & z>0,\,\, z'<0 \\
             \frac{4\pi i k_{\shortparallel} c^2}{\omega^2 \epsilon^d}\frac{k_d}{k_m \epsilon^d -k_d \epsilon^m}\,e^{i k_m z+i k_d z'}, & z<0,\,\,z'>0 \\
             \frac{2\pi i k_{\shortparallel} c^2}{\omega^2 \epsilon^d}\big[\frac{k_m \epsilon^d +k_d \epsilon^m}{k_m \epsilon^d -k_d \epsilon^m}\,
e^{i k_d (z+z')}+e^{i k_d |z-z'|}\,sgn(z-z')\big], & z>0,\,\, z'>0
           \end{array}
         \right.
\end{equation}
where
\begin{eqnarray}
  k_m &=& -\sqrt{\epsilon^m (\omega)\frac{\omega^2}{c^2}-k^2_{\shortparallel}},\nonumber \\
  k_d &=& \sqrt{\epsilon^d (\omega)\frac{\omega^2}{c^2}-k^2_{\shortparallel}},\nonumber\\
  \epsilon^m &=& \epsilon^m (\omega),\nonumber \\
  \epsilon^d &=& \epsilon^d (\omega).\nonumber
\end{eqnarray}
\section{}
%%%%%%%%%%%%%%%%%%%%%%%%%%%%%%%%%%%%%%%%%%%%%%%%%%%%%%%%%%%%%%%%%%%%%%%%%%%%%%%%%%%%%%%%%%%%%%%%%%%%%%%%%%%%%%%%%%%%%%%%%%%%%%%%%%%%%%%%%%%%%%%%%%%%%%%%%%%%%%%%
In the following the remaining components of $D_{spp}$ are given. The details of calculations are expressed in section III
\begin{eqnarray}\label{DXX}
\small&&\mathbf{D^{z,z'>0}_{xx,spp}}(\mathbf{X},\mathbf{X}';\omega)=\frac{-i|k_{spp}|}{4}\sqrt{\frac{\varepsilon_{d}}{-\varepsilon_{m}}}
\frac{\varepsilon_{d}\varepsilon_{m}^2}{(\varepsilon_{d}+\varepsilon_{m})(\varepsilon_{d}^2-\varepsilon_{m}^2)}
\,e^{-\sqrt{\frac{\varepsilon_{d}}{-\varepsilon_{m}}}|k_{spp}|(z+z')}\nonumber\\
&&(\frac{2(x-x')^2\,J_{0}(|k_{spp}||\vec{r}_{\|}-\vec{r}'_{\|}|)}{|k_{spp}||\vec{r}_{\|}-\vec{r}'_{\|}|^3}-\frac{2\,J_{1}(|k_{spp}|| \vec{r}_{\|}-\vec{r}'_{\|}|}{|k_{spp}|| \vec{r}_{\|}-\vec{r}'_{\|}|}-\frac{2(x-x')^2\,(J_{0}(|k_{spp}|| \vec{r}_{\|}-\vec{r}'_{\|}|)-J_{2}(|k_{spp}|| \vec{r}_{\|}-\vec{r}'_{\|}|))}{| \vec{r}_{\|}-\vec{r}'_{\|}|^2}),\nonumber\\
\end{eqnarray}
\begin{eqnarray}\label{DYY}
\small&&\mathbf{D^{z,z'>0}_{yy,spp}}(\mathbf{X},\mathbf{X}';\omega)=\frac{-i|k_{spp}|}{4}\sqrt{\frac{\varepsilon_{d}}{-\varepsilon_{m}}}
\frac{\varepsilon_{d}\varepsilon_{m}^2}{(\varepsilon_{d}+\varepsilon_{m})(\varepsilon_{d}^2-\varepsilon_{m}^2)}
\,e^{-\sqrt{\frac{\varepsilon_{d}}{-\varepsilon_{m}}}|k_{spp}|(z+z')}\nonumber\\
&&(\frac{2(y-y')^2\,J_{0}(|k_{spp}||\vec{r}_{\|}-\vec{r}'_{\|}|)}{|k_{spp}||\vec{r}_{\|}-\vec{r}'_{\|}|^3}-\frac{2\,J_{1}(|k_{spp}|| \vec{r}_{\|}-\vec{r}'_{\|}|}{|k_{spp}|| \vec{r}_{\|}-\vec{r}'_{\|}|}-\frac{2(y-y')^2\,(J_{0}(|k_{spp}|| \vec{r}_{\|}-\vec{r}'_{\|}|)-J_{2}(|k_{spp}|| \vec{r}_{\|}-\vec{r}'_{\|}|))}{| \vec{r}_{\|}-\vec{r}'_{\|}|^2}),\nonumber\\
\end{eqnarray}
\begin{eqnarray}\label{DZX}
\small \mathbf{D^{z,z'>0}_{zx,spp}}(\mathbf{X},\mathbf{X}';\omega)=\frac{-i|k_{spp}|}{2}
\frac{\varepsilon_{d}\varepsilon_{m}^2}{(\varepsilon_{d}+\varepsilon_{m})(\varepsilon_{d}^2-\varepsilon_{m}^2)}
\,e^{-\sqrt{\frac{\varepsilon_{d}}{-\varepsilon_{m}}}|k_{spp}|(z+z')}
\frac{(x-x')}{|\vec{r}_{\|}-\vec{r}'_{\|}|}\,J_{1}(|k_{spp}|| \vec{r}_{\|}-\vec{r}'_{\|}|),\nonumber\\
\end{eqnarray}
\begin{eqnarray}\label{DZY}
\small\mathbf{D^{z,z'>0}_{zy,spp}}(\mathbf{X},\mathbf{X}';\omega)=\frac{-i|k_{spp}|}{2}
\frac{\varepsilon_{d}\varepsilon_{m}^2}{(\varepsilon_{d}+\varepsilon_{m})(\varepsilon_{d}^2-\varepsilon_{m}^2)}
\,e^{-\sqrt{\frac{\varepsilon_{d}}{-\varepsilon_{m}}}|k_{spp}|(z+z')}
\frac{(y-y')}{|\vec{r}_{\|}-\vec{r}'_{\|}|}\,J_{1}(|k_{spp}|| \vec{r}_{\|}-\vec{r}'_{\|}|),\nonumber\\
\end{eqnarray}
\begin{eqnarray}
  \mid \vec{r}_{\|}-\vec{r}'_{\|}\mid=\sqrt{(x-x')^2+(y-y')^2}.
\end{eqnarray}
\section{}
%%%%%%%%%%%%%%%%%%%%%%%%%%%%%%%%%%%%%%%%%%%%%%%%%%%%%%%%%%%%%%%%%%%%%%%%%%%%%%%%%%%%%%%%%%%%%%%%%%%%%%%%%%%%%%%%%%%%%%%%%%%%%%%%%%%%%%%%%%%%%%%%%%%%%%%%%%%%%%%%
The free space Green's function is
\begin{equation}\label{G0}
\mathbf{G}_{0}(\vec{r},\vec{r}')=[\mathbf{I}+\frac{1}{k^2}\vec{\nabla}\vec{\nabla}]\,\frac{e^{ikR}}{4\pi R},
\end{equation}
\begin{equation}\label{G0}
\mathbf{G}_{0}(\vec{r},\vec{r}')=[(\frac{3}{k^2R^2}-\frac{3i}{kR}-1)\,\widehat{R}\widehat{R}+(1+\frac{i}{kR}-\frac{1}{k^2R^2})\,\mathbf{I}]\,
\frac{e^{ikR}}{4\pi R},
\end{equation}
where
\begin{equation}\label{G0}
R=\mid \vec{r} - \vec{r}'\mid,
\end{equation}
and
\begin{equation}\label{G0}
\widehat{R}=\frac{\vec{r} - \vec{r}'}{\mid \vec{r} - \vec{r}'\mid}.
\end{equation}
We made use of this function while calculating the electric field of a dipole in free space.
%%%%%%%%%%%%%%%%%%%%%%%%%%%%%%%%%%%%%%%%%%%%%%%%%%%%%%%%%%%%%%%%%%%%%%%%%%%%%%%%%%%%%%%%%%%%%%%%%%%%%%%%%%%%%%%%%%%%%%%%%%%%%%%%%%%%%%%%%%%%%%%%%%%%%%%%%%%%%%%%
\section{}
%%%%%%%%%%%%%%%%%%%%%%%%%%%%%%%%%%%%%%%%%%%%%%%%%%%%%%%%%%%%%%%%%%%%%%%%%%%%%%%%%%%%%%%%%%%%%%%%%%%%%%%%%%%%%%%%%%%%%%%%%%%%%%%%%%%%%%%%%%%%%%%%%%%%%%%%%%%%%%%%
\begin{equation}\label{EZ.Z}
E_{z,spp}(r,t)=\frac{\mu_{0}ql\Omega^2}{4}
|k_{spp}|\frac{\varepsilon_{d}\varepsilon_{m}^3}{\sqrt{\varepsilon_{d}(-\varepsilon_{m})}(\varepsilon_{d}+\varepsilon_{m})(\varepsilon_{d}^2-\varepsilon_{m}^2)}
\,J_{0}(|k_{spp}|\sqrt{x^2+y^2})\,e^{-\sqrt{\frac{\varepsilon_{_{d}}}{-\varepsilon_{m}}}|k_{spp}|(z+h)}\,\cos \Omega t,
\end{equation}
\begin{equation}\label{EX.Z}
E_{x,spp}(r,t)=\frac{-\mu_{0}ql\Omega^2}{4}
|k_{spp}|\frac{\varepsilon_{d}\varepsilon_{m}^2}{(\varepsilon_{d}+\varepsilon_{m})(\varepsilon_{d}^2-\varepsilon_{m}^2)}
\,\frac{x}{\sqrt{x^2+y^2}}\,J_{1}(|k_{spp}|\sqrt{x^2+y^2})\,e^{-\sqrt{\frac{\varepsilon_{_{d}}}{-\varepsilon_{m}}}|k_{spp}|(z+h)}\,\cos \Omega t,
\end{equation}
\begin{equation}\label{EY.Z}
E_{y,spp}(r,t)=\frac{-\mu_{0}ql\Omega^2}{4}
|k_{spp}|\frac{\varepsilon_{d}\varepsilon_{m}^2}{(\varepsilon_{d}+\varepsilon_{m})(\varepsilon_{d}^2-\varepsilon_{m}^2)}
\,\frac{y}{\sqrt{x^2+y^2}}\,J_{1}(|k_{spp}|\sqrt{x^2+y^2})\,e^{-\sqrt{\frac{\varepsilon_{_{d}}}{-\varepsilon_{m}}}|k_{spp}|(z+h)}\,\cos \Omega t,
\end{equation}
\begin{eqnarray}\label{EX.X}
E_{x,spp}(r,t)&=&\frac{\mu_{0}ql\Omega^2}{8}
|k_{spp}|\sqrt{\frac{\varepsilon_{d}}{-\varepsilon_{m}}}\frac{\varepsilon_{d}\varepsilon_{m}^2}{(\varepsilon_{d}+\varepsilon_{m})(\varepsilon_{d}^2-
\varepsilon_{m}^2)}
\,e^{-\sqrt{\frac{\varepsilon_{_{d}}}{-\varepsilon_{m}}}|k_{spp}|(z+h)}\,\cos \Omega t\nonumber\\
&(&\frac{2x^2\,J_{0}(|k_{spp}|\sqrt{x^2+y^2})}{|k_{spp}|(\sqrt{x^2+y^2})^3}-\frac{2\,J_{1}(|k_{spp}|\sqrt{x^2+y^2})}{|k_{spp}|\sqrt{x^2+y^2}}-
\frac{2x^2\,(J_{0}(|k_{spp}|\sqrt{x^2+y^2})-J_{2}(|k_{spp}|\sqrt{x^2+y^2}))}{(\sqrt{x^2+y^2})^2}),\nonumber\\
\end{eqnarray}
\begin{eqnarray}\label{EY.X}
E_{y,spp}(r,t)&=&\frac{\mu_{0}ql\Omega^2}{8}
|k_{spp}|\sqrt{\frac{\varepsilon_{d}}{-\varepsilon_{m}}}\frac{\varepsilon_{d}\varepsilon_{m}^2}{(\varepsilon_{d}+\varepsilon_{m})(\varepsilon_{d}^2-
\varepsilon_{m}^2)}
\,e^{-\sqrt{\frac{\varepsilon_{_{d}}}{-\varepsilon_{m}}}|k_{spp}|(z+h)}\,\cos \Omega t\nonumber\\
&(&\frac{2y^2\,J_{0}(|k_{spp}|\sqrt{x^2+y^2})}{|k_{spp}|(\sqrt{x^2+y^2})^3}-\frac{2\,J_{1}(|k_{spp}|\sqrt{x^2+y^2})}{|k_{spp}|\sqrt{x^2+y^2}}-
\frac{2y^2\,(J_{0}(|k_{spp}|\sqrt{x^2+y^2})-J_{2}(|k_{spp}|\sqrt{x^2+y^2}))}{(\sqrt{x^2+y^2})^2}),\nonumber\\
\end{eqnarray}
\begin{equation}\label{EZ.X}
E_{z,spp}(r,t)=\frac{\mu_{0}ql\Omega^2}{4}
|k_{spp}|\frac{\varepsilon_{d}\varepsilon_{m}^2}{(\varepsilon_{d}+\varepsilon_{m})(\varepsilon_{d}^2-\varepsilon_{m}^2)}
\,\frac{x}{\sqrt{x^2+y^2}}\,J_{1}(|k_{spp}|\sqrt{x^2+y^2})\,e^{-\sqrt{\frac{\varepsilon_{_{d}}}{-\varepsilon_{m}}}|k_{spp}|(z+h)}\,\cos \Omega t.
\end{equation}
%%%%%%%%%%%%%%%%%%%%%%%%%%%%%%%%%%%%%%%%%%%%%%%%%%%%%%%%%%%%%%%%%%%%%%%%%%%%%%%%%%%%%%%%%%%%%%%%%%%%%%%%%%%%%%%%%%%%%%%%%%%%%%%%%%%%%%%%%%%%%%%%%%%%%%%%%%%%%%%%

\end{document}